# Automatic Classification of Alzheimer's Disease using brain MRI data and deep Convolutional Neural Networks

**Zahraa Sh. Aaraji, Hawraa H. Abbas**

**zahraa.sh@s.uokerbala.edu.iq, hawraa.h@uokerbala.edu.iq**

Collage of engineering , Electrical and Electronic department, University of Kerbala, Kerbala, Iraq

*Abstract*:

**Alzheimer's disease (AD) is one of the most common public health issues the world is facing today. This disease has a high prevalence primarily in the elderly accompanying memory loss and cognitive decline. AD detection is a challenging task which many authors have developed numerous computerized automatic diagnosis systems utilizing neuroimaging and other clinical data. MRI scans provide high-intensity visible features, making these scans the most widely used brain imaging technique. In recent years deep learning has achieved leading success in medical image analysis. But a relatively little investigation has been done to apply deep learning techniques for the brain MRI classification. This paper explores the construction of several deep learning architectures evaluated on brain MRI images and segmented images. The idea behind segmented images investigates the influence of image segmentation step on deep learning classification. The image processing presented a pipeline consisting of pre-processing to enhance the MRI scans and post-processing consisting of a segmentation method for segmenting the brain tissues. The results show that the processed images achieved a better accuracy in the binary classification of AD vs. CN (Cognitively Normal) across four different architectures. ResNet architecture resulted in the highest prediction accuracy amongst the other architectures (90.83% for the original brain images and 93.50% for the processed images).**

# 1 INTRODUCTION:

AD causes many abnormalities that develop structural changes in different brain regions. Such structural changes appear in all people as they age but are less evident. AD is a chronic condition that causes the death of the brain cells leading to cognitive impairment. Cognitive mental troubles such as forgetfulness and confusion are some of the most prominent symptoms of Alzheimer's patients [1].

The primary cause of AD is the accumulation of two types of brain proteins, Beta Amyloids (Aβ) and Tau Tangles. These accumulations disrupt the normal functioning of the neurons and trigger a series of toxic events. In this case, neurons get damaged, lose their connections to the surrounding neurons, and eventually die [2]. According to scientists, protein accumulation seems to be related to a combination of environmental and genetic factors. The most common risk factor associated with developing AD is aging. Other environmental risk factors include smoking, strokes, heart diseases, depression, arthritis, and diabetes.

In the late stages of Alzheimer's, damages become diffuse, and the brain shrinks significantly. Because beta amyloids grow gradually over time, it will take over (10) years before a patient starts noticing any clear signs of the disease. In brain scans, one can see the widespread damage inside the brain resulting from the dead neurons.

AD is considered the fifth leading cause of death in adults older than 65 years. An estimated 6.2 million Americans 65 years or older are having their life with AD in 2021. This number is projected to increase to 13.8 million by 2050. The estimated total healthcare costs for the treatment of AD in 2021 is $355 billion, and these costs are expected to rise to more than $1.1 trillion in 2050 [3], [4].

The costs for AD patients' care are expected to rise dramatically. Therefore, early detection of individuals at risk has become a critical issue. Early detection helps to slow down the disease and reduce the costs of treatment because it enables people with dementia and their families to better prepare for the progression of the disease. A study on more than 3000 cases defined a score for a healthy lifestyle including nonsmoking, physical activity, low alcohol consumption, high quality dietary, and engagement in cognitive activities. The study concluded that a high score healthy lifestyle is associated with a lower risk factor of AD [5].

Clinically, numerous medical imaging techniques like Magnetic Resonance Imaging (MRI), Positron Emission Tomography (PET), and Computed Tomography (CT), coupled with advanced computational machine learning methods, have led to accurate prediction of the disease presence. MRI scanners create a detailed 3D image of the brain employing magnetic fields and radio waves. Brain MR Images can reveal the structural atrophic changes in the brain. Figure (1) shows the three MRI planes (taken from the ADNI datasets).

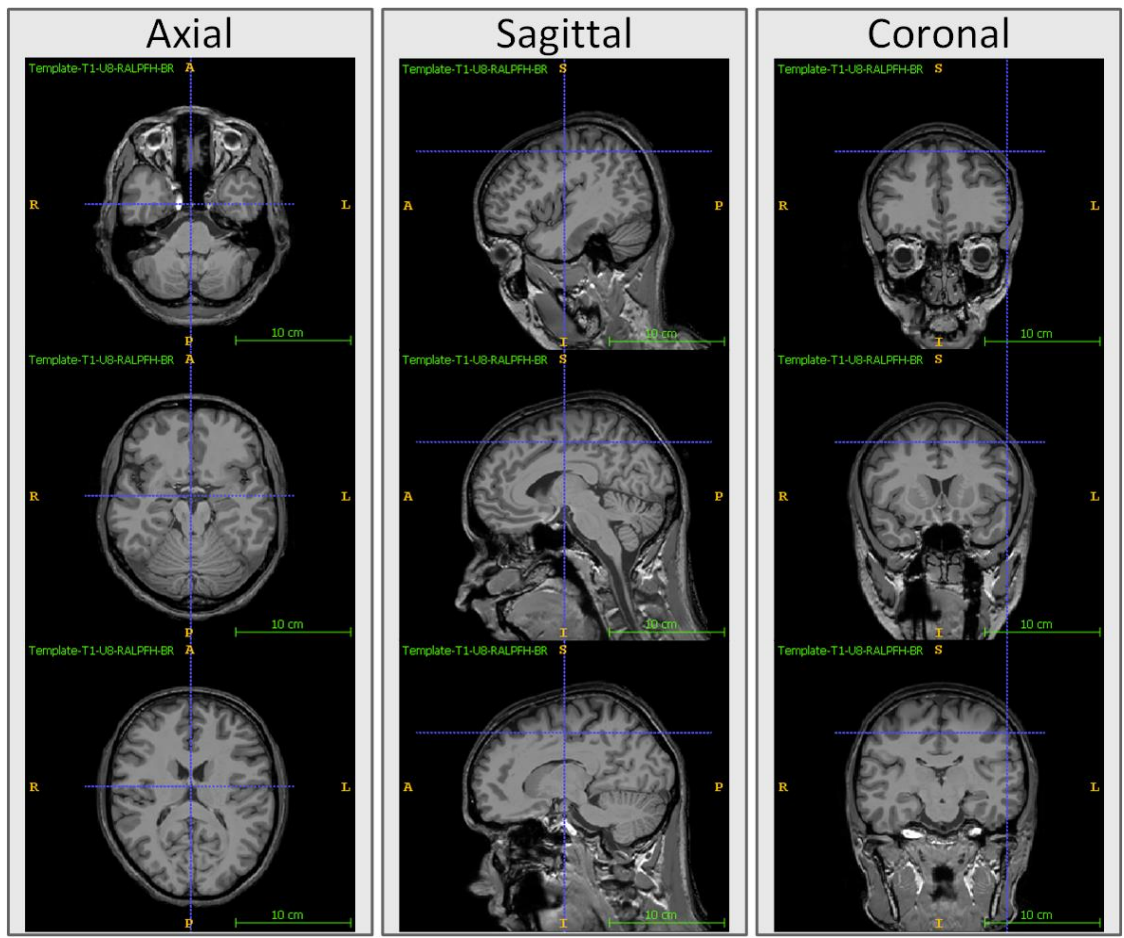

*Figure 1: Axial, Coronal and Sagittal views of a 3D T1-weighted MR image.*

Machine learning studies using neuroimaging data helped a lot in the automated brain MRI analysis [6], [7]. Deep Learning (DL) has emerged as one of the most promising machine learning tools for AD diagnosis. Discriminative features can be extracted automatically from raw data using DL models. The most advanced DL architectures are designed to work with real-world images for image segmentation, regression, and classification. These models require a large amount of training data such as brain MR images to learn the patterns and features embedded in these images. The benefit of DL models is that learned features are directly extracted from input images, eliminating the need for manually generated AD features. [8], [9].

This paper investigates four different DL architectures in diagnosing AD based on T1-weighted structural MRI data. This paper also evaluates the effectiveness of the image segmentation on The DL model for feature extraction and classification. Hence, this paper compares brain images (original images) against tissue-segmented brain images.

Each MR image undergoes three pre-processing steps: intensity normalization, spatial registration, brain extraction. Then, the brain's three tissues: White Matter (WM), Gray Matter (GM), and Cerebrospinal Fluid (CSF), are segmented using the Hidden Markov Random Field (HMRF) segmentation technique. The segmented tissue images are fed into the Deep Convolutional Neural Networks for the feature extraction task.

The experimental results show that the tissue-segmented images have achieved better accuracy than the original images; this proves that the tissue-segmented images are more efficient in AD diagnosis. The main contribution of this paper can be summarized as follow:

- Exploit and train four advanced DL architectures not pre-trained on the ImageNet dataset.
- Training the DL models from scratch reduces the complexity of the architectures and offers exclusivity to brain patterns.
- Utilizing image segmentation reduces the complexity of the MRI data and facilitates the DL models training.

This paper is organized as follows: Section 2 offers a brief review of different classification methods that have been reported in the literature for this problem. The data used in this paper are described in Section 3, the proposed approach is introduced in Section 4. In Section 5, experimental results are provided, the discussion is provided in section 6, and finally, in section 7, the conclusion and future work is presented. Figure (2) illustrates the block diagram of the proposed methodology.

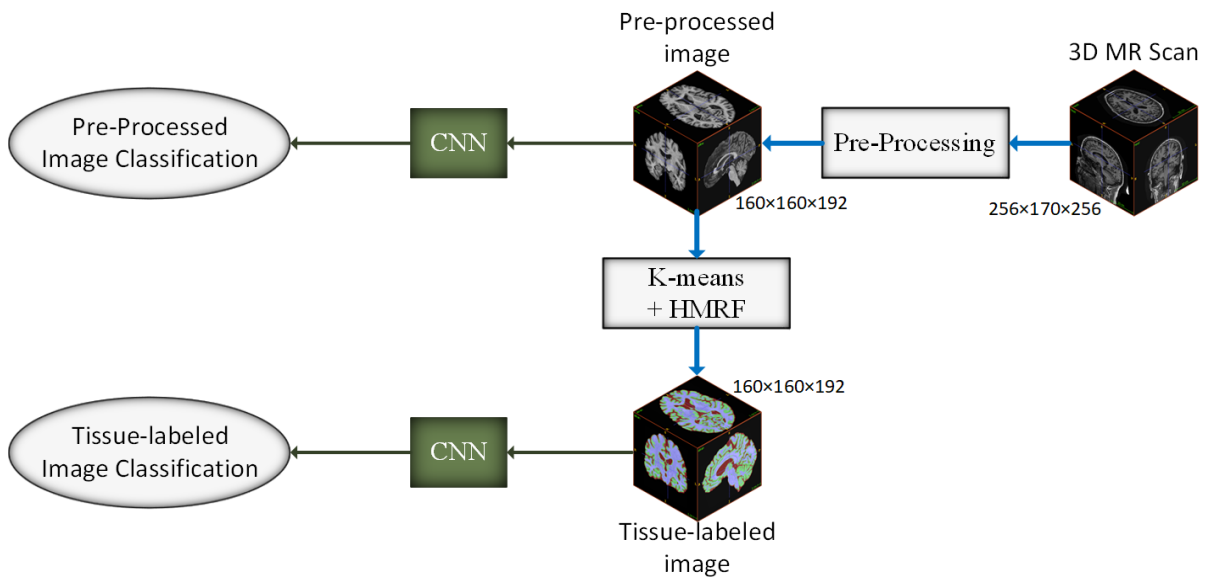

*Figure 2: Block diagram of this paper investigations*

# 2 RELATED WORKS:

DL is a relatively new ML methodology that has recently made significant advancements in medical image processing and recognition, including image segmentation, object detection, features extraction, and classification. CNNs have attracted a lot of attention as the most widely used DL architectures because of their high success in image classification. Unlike traditional ML algorithms, DL methods automatically learn unique characteristics from raw data because of their hierarchical and layer-wise structure. DL architectures allow for the automatic abstraction of low- to high-level latent feature representations (e.g., dots, lines, and edges for low-level features, and circles, cubes, and many other shapes and objects for high-level features). As a result, it is right to assume that DL relies less on the human and requires less prior knowledge of other complex procedures [10]–[12].

Image classification and its application have made considerable progress in the last years. The academic circles have made great efforts to design various CNN models that have achieved high accuracy and even exceeded the human recognition ability. One example of this classification function comes from a breast cancer diagnosis [13]–[15]. Breast cancer diagnoses typically include an assessment by pathologists of a tissue sample, in which doctors look for certain defined features that indicate the presence of the disease. A deep learning system trained on tissue images can achieve higher accuracy than pathologists by finding and utilizing the embedded patterns of the medical image. In doing so, such diagnostic tools were able to help doctors more accurately assess a patient's prognosis.

Another example comes from the diagnosis of brain tumors [16]. With the application of DL algorithms on the MRI scans, the prediction of the brain tumor presence is very fast with higher accuracy. This prediction helps the radiologist in making quick decisions. Additionally, DL has also contributed to other diseases such as lungs cancer and Alzheimer's disease [17], [18]. This paper will focus on CNN architectures exploited for AD diagnosis using structural MRI scans.

(Suk and Shen [19], 2013) were the first to use a DNN architecture that consists of an input layer, three hidden layers, and an output layer. They segmented MRI images into gray matter, white matter, and CSF, and then parcellated them into (93) ROIs. They employed the DNN architecture to learn the compressed features of the computed GM volumes from the segmented ROIs. Following feature extraction and selection, they implemented an SVM classifier to evaluate the learned features for subject classification. While Suk and Shen did use a DL architecture, they did not exploit the DNN architecture capability to analyze the MR images directly but a mere feature selection method.

The increase of studies in DL related to AD diagnosis relates to its ease of use. In traditional ML techniques, obtaining well-defined features is necessary, but these features become increasingly hard to find as the complexity of the data increases. DL simplicity comes from the fact that the user does not have to decide which features are efficient; the DL hidden layers decide.

In machine learning techniques the Pre-processing, feature extraction and selection requires a great deal of reliability by users. Since it significantly affects the classification performance. DL approaches have been applied to AD diagnostic classification using original neuroimaging data without any feature selection procedures. Pre-processing is optional when using DL architecture, although it reduces DL complexity. Using whole- or image patches of the brain necessitates an extremely high number of parameters making DL approaches demand very high computational power. The advancement of computational power and the rise of Graphical Processing Units (GPU) for parallel computing made training times of deeper DL architectures much lower. This advancement facilitated the possibility of applying brain MR images as an input to DL networks for feature extraction, selection, and classification.

(Puente-Castro et al. [20], 2020) implemented the trained 2D ResNet to extract features from the sagittal view of MR images. These features are inputted to a final SVM classifier to diagnose AD. (Valliani et al. [21], 2017) Utilized the pre-trained ResNet-18 network for the classification. Input data were a single image of the median axial slice skull-stripped and registered. The last two fully-connected layers were trained from scratch to output both multi-class and (AD vs. CN) predictions. (Pan et al. [22], 2020) designed (123) homogeneous CNNs with (6) convolution layers and (2) fully connected layers for a set of slices consisting of (40) sagittal slices, (50) coronal slices, and (33) axial slices. The outputs from the multiple trained CNN classifiers were then combined to generate the final classifier ensemble to predict the classification results. The proposed method enabled the authors to inspect most brain regions that contribute to the discrimination of (AD vs. CN) subjects.

(Islam et al. [23],2018) used an ensemble of CNNs with Dense layers for all three views: Sagittal, Coronal, and axial, to identify different stages of AD. Results show that pre-trained networks surpass shallow networks, but 2D pre-trained networks lose most of the information embedded in the 3D MR images when sliced and analyzed by 2D convolutional layers. Using all three views of MR images reduces this information loss to a certain limit.

In recent researches, authors focused on 3D networks to address the problem of insufficient information in the 2D slice-level approach. Despite their higher computational cost, these models are better when extracting discriminative features from the 3D brain in MR images.

(Yagis et al. [24], 2020) designed a Deep 3D CNN model similar to VGG-16 architecture for binary classification (AD vs. CN). His network accepts images without any pre-processing steps to output the disease state. This approach was evaluated on two datasets, ADNI and OSAIS, using T1-weighted structural brain MR images. (Qiu et al. [25], 2018) employed the pre-trained VGGNet-11 for the AD vs. CN classification. Input data were only (3) slices from the 3D MR images. They froze all convolution layers and kept the last fully-connected layers for the training. They trained (3) models for each slice, and the output prediction was fused using a series of voting approaches. Their result confirms that using more image slices increases the prediction accuracy.

The limitation of these studies employed one or more 2D slices of the 3D brain MRI. They achieved low classification accuracy because of information loss when neglecting other brain regions not included in the chosen 2D slices.

(Li et al. [26], 2018) proposed a classification method based on multiple cluster dense CNN (DenseNet) to learn the various local features of 3D patches extracted from MRI brain images and clustered using the K-means algorithm. Local patch-level features resulting from each patch for the same MR image are ensembled to form the final global classification results. This method was evaluated on AD vs. CN subjects downloaded from the ADNI dataset. (Cheng et al. [27], 2017) proposed a similar multiple deep 3D CNN to Li et al. that differs only in the CNN architecture. Cheng et al. Extracted local features from the MRI brain patches and fused them to form the final diagnosis.

The benefit of using small image patches as CNN inputs reduces the computational power requirements to train such CNN networks. That is because an input image with fewer pixels requires fewer convolutions, pooling, and other operations. Fewer operations mean that the network training requires less memory size to hold the activation maps of a single iteration. Patch-level networks' drawback is that they lack the continuity of the whole image resulting in discontinuous patterns in-between neighboring patches.

(Korolev et al. [28] ,2017) proposed two deep 3D CNN architectures to classify full brain MRI scans. The first architecture is similar to the VGGNet, and the second architecture matches the ResNet architecture. Both CNNs implement 3D convolution and pooling layers instead of 2D ones. Both networks showed similar results for the AD classification. (Zhang et al. [29] ,2021) used 3D ResNet architecture for classifying (AD vs. CN). They utilized the Self-Attention residual mechanism to capture long-range dependencies and reduce computational inefficiency due to repeated convolutional operations. They also implemented gradient-based localization class activation mapping (Grad-CAM) to visualize and explain the prediction of AD. Results have shown that the classification performance of models with the self-attention mechanism is significantly higher than models without it.

The studies utilized whole-brain MR images to automate feature extraction and classification of AD. However, these methods have several limitations: they are computationally intensive because of CNN complexity, in addition to they are prone to over-fitting due to high dimensionality.

In this paper, we investigate the influence of image processing on DL classification performance. Thus, we study the impact of reducing data complexity using image segmentation on the final classification results of AD diagnosis. We redesigned (4) distinct DL architectures to evaluate the segmented images compared to regular pre-processed MR images.

# 3 MATERIALS:

Alzheimer Disease Neuroimaging Initiative (ADNI) database (adni.loni.usc.edu) was the first source of brain MR images and the clinical data used in evaluating the proposed method. ADNI has been running since 2004, and funds will last until 2021. The primary goal of ADNI is to test whether serial MRI, PET scans, other biological markers, clinical and neuropsychological assessment can track and diagnose the early stages of AD progression. The dataset also contains metadata for each brain scan, including gender, age, education, and diagnosis. Determination of sensitive and specific markers of early AD progression intends to aid researchers and clinicians in developing new treatments, monitoring their effectiveness, and lessening the time and cost of clinical trials.

This paper uses Three-dimensional T1-weighted structural MRI from baseline groups of ADNI1. ADNI1 provides 435 subjects split into two groups: AD group refers to the patients with Alzheimer's diagnosis (n = 200), CN group refers to normal cognitive status subjects that show no sign of AD (n =200). Only (200) CN subjects are chosen from the CN group to sustain classes balance and prevent classification bias. The standards for discarding (35) CN subjects are random and based on (5) age groups to ensure age variation among the chosen (200) CN group. . This data are pass through several steps detailed in methodology to preparation for AD classification

# 4 METHODOLOGY:

This section describes the proposed classification pipeline used for the MRI classification. The MRI classification pipeline includes four primary steps: pre-processing, post-processing, feature extraction, and feature classification. This work was implemented using the latest MATLAB version (2021a) on a single PC with an Intel 10th Gen Core i7 processor, Nvidia RTX 2070, and 32 Gigabyte system RAM.

## 4.1 Pre-Processing Stage:

MR images downloaded from the ADNI must go through (3) pre-processing steps before further analyzing these images. The pre-processing steps enhance the quality of the original MR images by removing unneeded objects, normalizing intensity variations, and aligning images to template space. Pre-processing ensures that all MR images are standardized and normalized to reduce the complexity of the following processes. The MR image pre-processing consists of: Brain extraction (Skull-stripping), Intensity normalization, and Image Registration.

### 4.1.1 Brain Extraction:

Brain extraction is substantial to the proceeding analysis steps such as image segmentation, registration, and classification. Removing unneeded objects from the head MRI scan ensures better classification performance. These undesired objects, like the skull, may affect pattern recognition and increase the complexity of input data. Since AD affects only the brain tissues, all non-brain tissues and objects in the MRI scan are undesired to the proceeding steps.

An accurate brain extraction method should exclude all non-brain tissues such as the skull, skin, eyes, and fat without removing any part of the brain. The importance of an accurate and robust brain extraction step reflects in the number of methods proposed to address the brain segmentation problem [30]. Most brain extraction methods work on T1-weighted MR images as it provides excellent contrast for the different brain tissues.

ADNI offers a brain tissue mask created using a semi-automated segmentation method. This mask provides accurate definitions of the GM and WM of the brain, but they exclude all interior and exterior CSF. Therefore, we used voxel-based morphological operations (dilation followed by erosion) to include the CSF. Image dilation operation followed by erosion operation is called a closing operation. Multiplication of the generated mask by its brain MR image results in skull removed image. . Figure (3) illustrates the morphological operations of the brain extraction process and the shape of the chosen structural element.

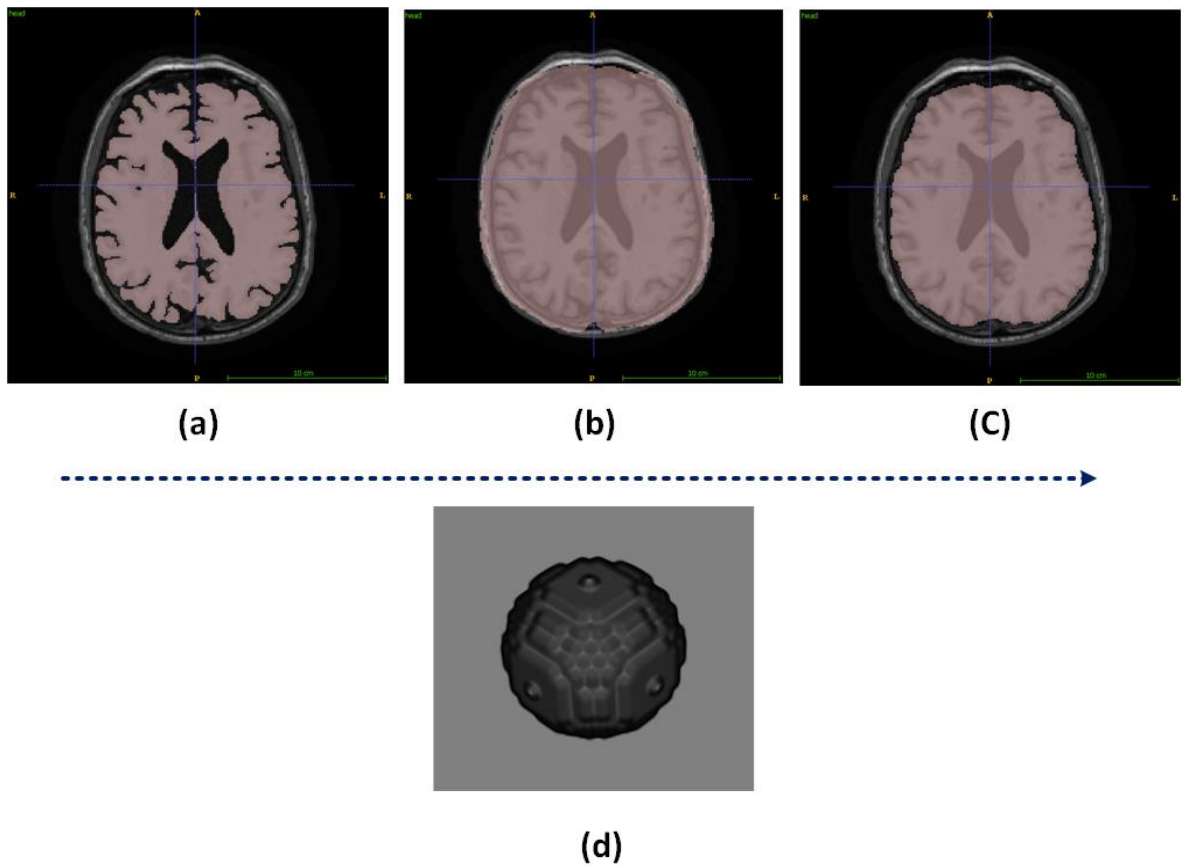

*Figure 3*: *illustrates the brain extraction process (a) Original mask (b) Dilated mask (c) Eroded mask (d)Spherical structural element.*

### 4.1.2    Intensity Normalization:

Different MRI scanners use different parameters that result in large intensity variations between MR images. Intensity variations reduce the quality and performance of the subsequent processes. And the post-processing stage would be negatively affected, lowering the final classification accuracies. The simple Intensity scaling technique is chosen based on three intensity regions:

- WM intensity (high-intensity region).
- GM intensity (medium-intensity region).
- CSF intensity (low-intensity region).

The $K$-means ($k = 3$) clustering algorithm can compute the (3) centroids that define the (3) intensity region. The intensity scale difference between the brain image and the ICBM152 template is calculated based on these (3) centroids. The software pipeline uses this scale difference to normalize each brain extracted image to the template.

### 4.1.3    Registration:

Another issue with using different MRI scanners is the spatial variation between head scans. Spatial variation means that the subjects' heads are not located in the matching coordinates or have the same scale and angle. However, to analyze the brain MR images, spatial correspondence of anatomy across different MR images is a must. It is only possible to compare structural variations for subjects with equivalent brains regions on a common spatial domain. Although many DL techniques are space invariant, spatial normalization reduces the number of patterns needed for the classification task.

Image registration is necessary to find correspondence between the brains, which a brain region in one MR image corresponds to the same brain region in all other MR images. The proposed pipeline adopts the affine registration for the brain alignment task of structural MRI scans to a common spatial domain [31]. Affine registration [32] is a rigid-body linear registration that tends to preserve parallelism, i.e., do not deform brain structures but rather manipulate linear parameters:

- Translation (head location).
- Rotation (head angle),
- Scaling (voxel size),
- Shearing (planar direction).

All subjects are registered to share a common space by the same ICBM152 2009a template. ICBM152 2009a template is a nonlinear symmetric T1 weighted brain MRI with a voxel size of $1 \times 1 \times 1$ mm3, Figure (4) represents the pre-processing steps applied to an example MR image. Notice how a defined axial slice presented different brain regions between a registered and non-registered MR image. Another advantageous outcome of the registration is the brain size normalization that highly affects the pattern recognition complexity.

Most downloaded MR images have a resolution of (256x160x256). Such high resolution contains more than (10) million voxels per single 3D image, and most voxels are background voxels that hold no information. This amount of data is very high, especially in the classification stage. To solve this, we cropped the template to the lowest possible resolution (160x160x192) with less than half the number of voxels (4915200). Removing most background voxels reduces the training time of the DL architecture to half, which usually takes hours or even days. Registering the downloaded MR images to the cropped template forces the template resolution on the MR images.

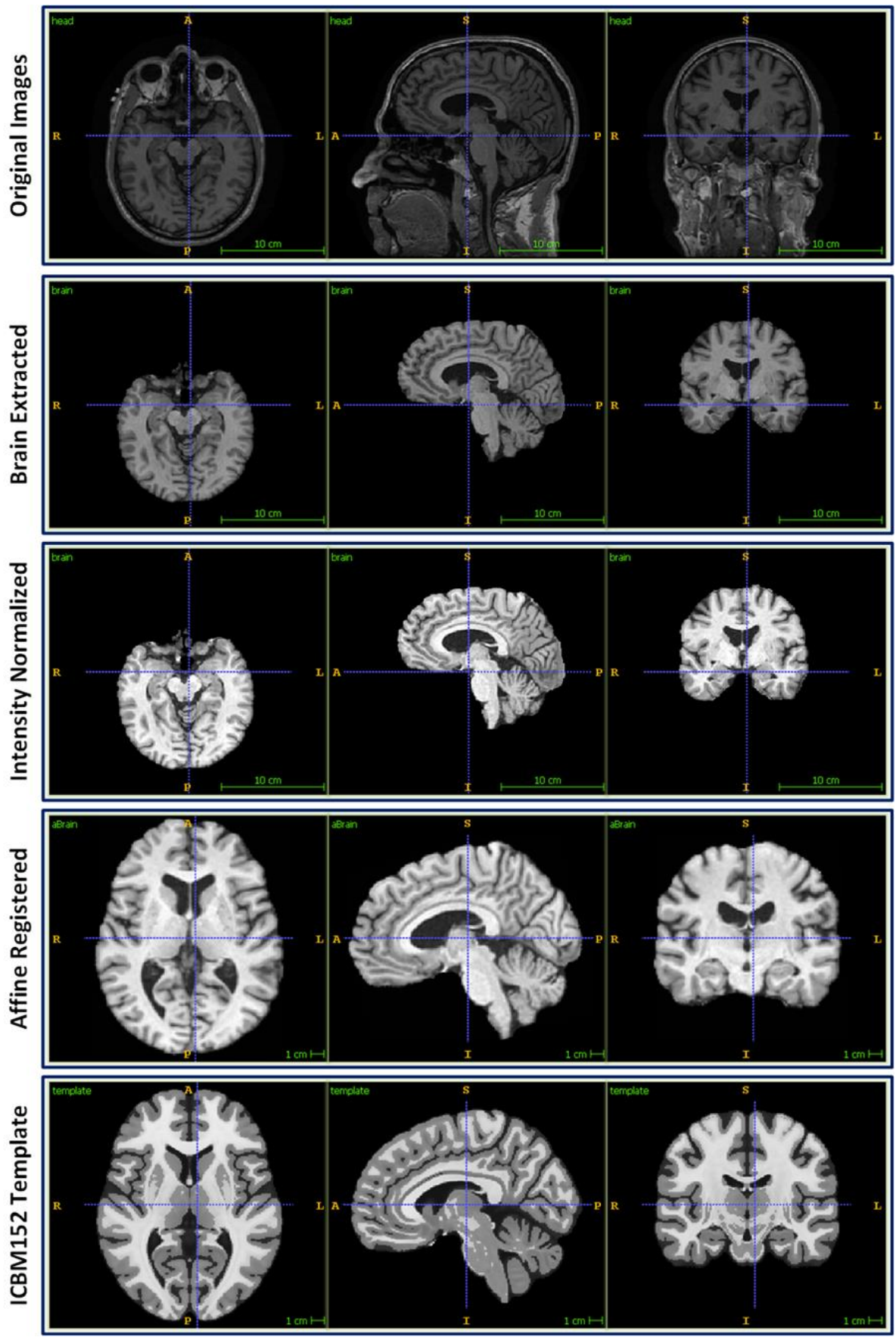

*Figure 4*: *Pre-processed MR images and ICBM152 template.*

## 4.2    Post-Processing Stage:

The post-processing stage performs brain segmentation on pre-processed MR images before passing them to the final stage (feature extraction and classification). This stage is needed to evaluate the development of image segmentation on classification performance. Image segmentation refers to dividing an image into multiple non-overlapping regions based on several characteristics like pixel intensity and contours.

MRI can capture the internal structures and anatomical components of the human brain. GM, WM, and CSF are the main components that represent the three anatomical tissues within the human brain. MRI scans can capture these brain tissues in a single 3D image. However, they vary in their representative intensities, and each tissue has an intensity region representing its voxels.

There are various segmentation techniques available. This paper evaluates several automated segmentation techniques. This process aims to reduce the complexity of the image by reducing the number of data bins, and thus proceeding analyses becomes Simpler. Segmenting an image into three brain tissues reduces voxels' intensity range to only three data bins or labels. These (3) bins represent the (3) tissue classes WM, GM, and CSF. The three MRI brain tissues contain features capable of discriminating AD from CN subjects.

In this paper, several segmentation methods are utilized and evaluated against each other to find an accurate segmentation method. Most segmentation techniques are statistical approaches that segment an image based on its voxels' intensity like $K$-means, Otsu, and Fuzzy C-means. The Genetic Algorithm (GA) and Particle Swarm Optimization (PSO) methods are compared against a statistical model called Hidden Markov Random Field (HMRF) for segmentation refinement. HMRF is a probabilistic model that can improve the segmentation by using other image characteristics like voxel neighborhood. Evaluation of these segmentation methods depends on the execution time and segmentation accuracy [33]–[39].

### 4.2.1      K-means Initial Segmentation:

The $K$-means clustering algorithm has been used widely because of its advantages like simplicity, efficiency, and ease of implementation. The $K$-means clustering method divides input data into ($K$) clusters by computing ($K$) centroids for each category iteratively. Then, $K$-means segments input data by grouping each input element to the closest centroid. In an image segmentation task that relies on pixel intensity, the centroid is the mean intensity of its clustered pixels.

The $K$-means clustering method forces pixels or voxels to belong exclusively to one class in each iteration. $K$-means algorithm may not give an optimum solution even after many iterations, mainly because of:

- Voxels' intensity may suffer from noise and variations.
- MRI scanners cannot capture a clear boundary between tissues.
- Voxels may represent more than one tissue (Partial Volume Effect).

Euclidean distance is one of the objective functions that $K$-means uses to measure how far a pixel is from its cluster centroid [33]. Because all brain MR images are intensity normalized, defining a starting point for the centroids is acceptable. The intensity ranges of the (3) tissue regions are known to initialize a centroid for each tissue class. For example, the intensity range for the CSF is in the first portion of the gray-scale MR image (e.g., 0 to 50). Using a pre-defined value for the $K$-centroids initialization leads to faster convergence.

$K$-means is also utilized in the pre-processing stage to compute the centroids of the brain regions. But no clustering is performed by the normalization process; the centroids are used only for calculating the scale difference between the brain image and the template.

### 4.2.2    Hidden Markov Random Field (HMRF):

HMRF is a probabilistic model that can refine a segmentation by adding more image characteristics. This concept can increase the segmentation accuracy of an intensity-based segmentation method like K-means. In this paper, voxels' neighborhood feature is embedded within the HMRF model to fine-tune the initial segmentation. Hence, the HMRF model can explain a voxel label by two probabilities: its intensity value and its neighboring voxels [35].

For example, a voxel located between two intensity regions cannot be classified correctly based on its intensity value only. The HMRF model will classify such voxels based on their neighboring voxels. The HMRF model accepts the intensity distances of all voxels from their cluster centroid and transforms these distances into probabilities. These probabilities explain how certain a voxel belongs to a cluster. Then, The HMRF model boosts these probabilities by how many neighbors of the same label this voxel had. Figure (5) illustrates the effect of the HMRF model and the voxel neighborhood system. Figure (6) shows 3views from a single subject MRI session along with segmentation processes.

HMRF implementation uses two procedures to compute the updated parameters that fine-tune the initial segmentation. The parameters of this process are the Gaussian parameters mean and variance for each tissue. The two procedures run sequentially in a defined number of iterations:

A. **Expectation Step**: This step uses the Iterated Conditional Modes Algorithm (ICM). ICM iteratively estimates the new labels for each voxel after inserting the voxel neighborhood term into the equation. The number of iterations in ICM represents how many times to update a voxel label based on the neighboring voxels. This ICM implementation uses a 3D $1^{st}$ order neighborhood system to compute the neighborhood probability for each voxel. At each iteration, the likelihood for each voxel equals the summation of the updated neighborhood term and the intensity term. The intensity term in ICM is the Probability Distribution Function (PDF) of the brain image voxels given the initial segmentation mean and variance. ICM passes the final likelihood of the last iteration to the second step.

$$Likelihood_i(x_j|l) = Neighborhood_i(x_j|l) + Intensity(x_j|l)$$

$Where: x_j = is\ the\ voxel\ at\ interest$
$\qquad\quad l = is\ the\ given\ tissue\ class$
$\qquad\quad i = is\ the\ ICM\ iteration$

In each iteration, ICM updates the voxels' labels based on the highest likelihood for the given tissue label. And when the voxels' labels get updated, each voxel gets new labeled neighboring voxels. This process changes the neighborhood term for that voxel for the next iteration. The intensity term depends solely on the voxel's intensity value and the Gaussian distribution parameters. Hence, this term remains the same at every iteration.

B. **Maximization Step**: The second step calculates the Gaussian distribution parameters of the final likelihood of the ICM algorithm. Updating the mean and variance of the Gaussian distribution result in a new intensity term. This step passes back this new intensity term to the ICM for the next iteration to get a new neighborhood-based likelihood term for each voxel given a label. These two iterative steps are called the Expectation-Maximization (EM). EM repeats for several iterations defined by the user.

The significance of the HMRF model is the explicit embedding of the 3D spatial relations between the image voxels into the intensity model. Consequently, a segmentation process powered by HMRF becomes more robust to noise, intensity variations, and voxels' PVE.

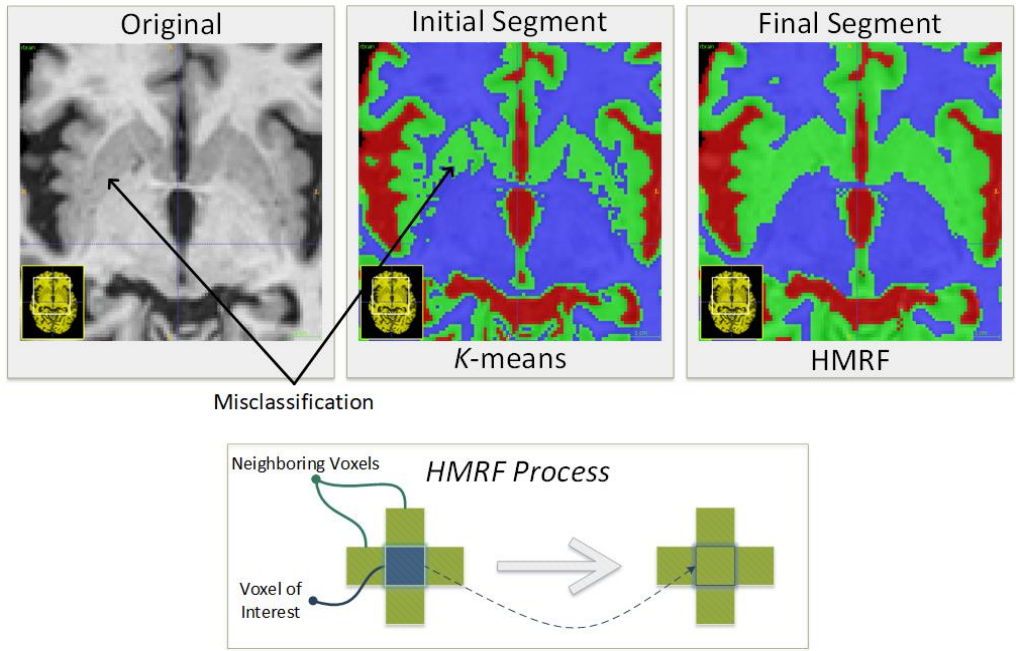

*Figure 5: shows the K-means followed by the Hidden Markov Random Field process.*

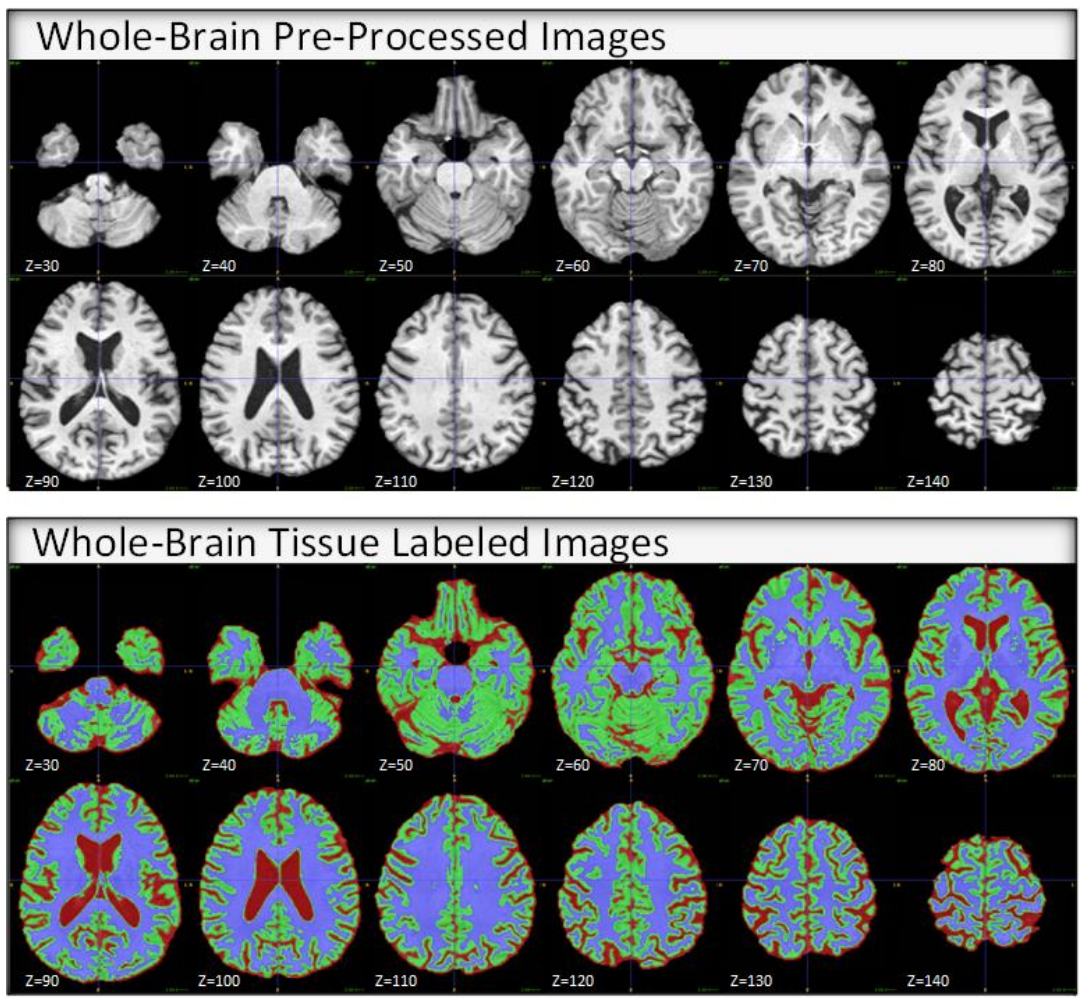

*Figure 4: shows (Axial, sagittal, and coronal) view respectively from left to right in a head scan from a single-subject MRI session along with segmentation processes.*

### 4.3    Feature extraction and Classification:

CNN architectures are the most common DL approach that achieved outstanding results in computer vision tasks such as image segmentation, feature extraction, and classification [9]–[11]. While traditional ML algorithms involve a human telling a machine what should be there, DL algorithms can automatically extract the desired features. CNNs consist of (3) layer types: Input layers, output layers, and hidden layers.

An input layer receives external data for pattern recognition tasks, whereas an output layer provides the solution to the problem. On the other hand, a hidden layer links between these input and output layers. Hidden layers are the essence of a DL architecture that process data entries and map the inputs to the outputs. CNNs Hidden layers typically have one or more convolution layers that do the convolution operation (between the filter and the input image), one or more pooling layers to perform Down-Sampling operations, non-linearity layers, and fully connected layers [40].

Convolution layers can detect many features within an image. These features might be one of many things like edges, lines, curves, transitions, etc. Activation layers apply non-linearity to the output of its previous layer, usually a convolution layer. This non-linearity helps the network to train faster by removing unnecessary values. The pooling operation reduces the dimensionality of the feature maps. CNNs can detect and extract complex patterns by doing a consecutive number of convolutions, activations, and pooling operations [41], [42]. The final output form of the CNN architecture before the fully-connected layers would be a vector of features in an image classification task. Fully-connected layers take full responsibility for transforming these features into a class probability that determines the category of the image.

### 4.3.1    CNN Architectures:

Choosing the best architecture suited for feature extraction and classification is a time-consuming task; it requires trial and error experiments to evaluate different architectures. Various CNN architectures exist; each has a different number of convolution, pooling, ReLU, fully connected, and custom layers. The most common architectures are detailed below. The figure below presents the block description for the follower block diagrams of the detailed CNN architectures.

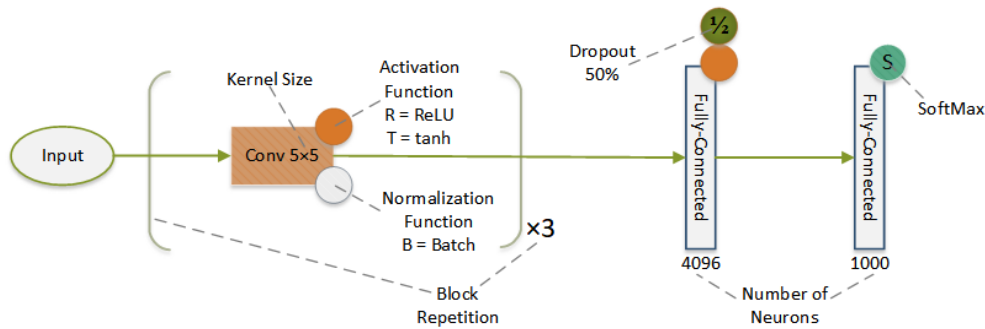

*Figure 7: Basic block diagram explanation.*

#### 4.3.1.1    AlexNet:

AlexNet architecture was developed by Krizhevsky et al. [43] in the year 2012. AlexNet has (61) million learnable parameters (filters) and was the best CNN architecture in its time. AlexNet has five convolution layers and three fully connected layers. Krizhevsky was the first to implement the ReLU activation function; this made his network much faster in the training phase. Figure (8) represents the adopted AlexNet architecture.

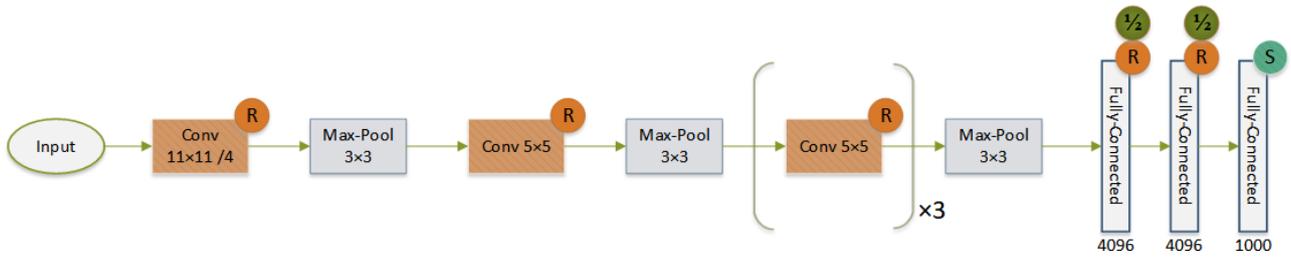

*Figure 8: AlexNet architecture.*

### 4.3.1.2    VGG-16:

VGG-16 was developed at Visual Geometry Group (VGG) by Karen Simonyan et al. [44] in 2014, and it used a deeper configuration of AlexNet. The architecture of VGG-16 generally consists of (13) convolution layers followed by ReLU activation and max-pooling layers in addition to (3) fully-connected layers. VGG-16 has (138) million learnable parameters, more than twice what AlexNet has. VGG-19 is a deeper model with more convolutions than VGG-16. Figure (9) illustrates the adopted VGG-16.

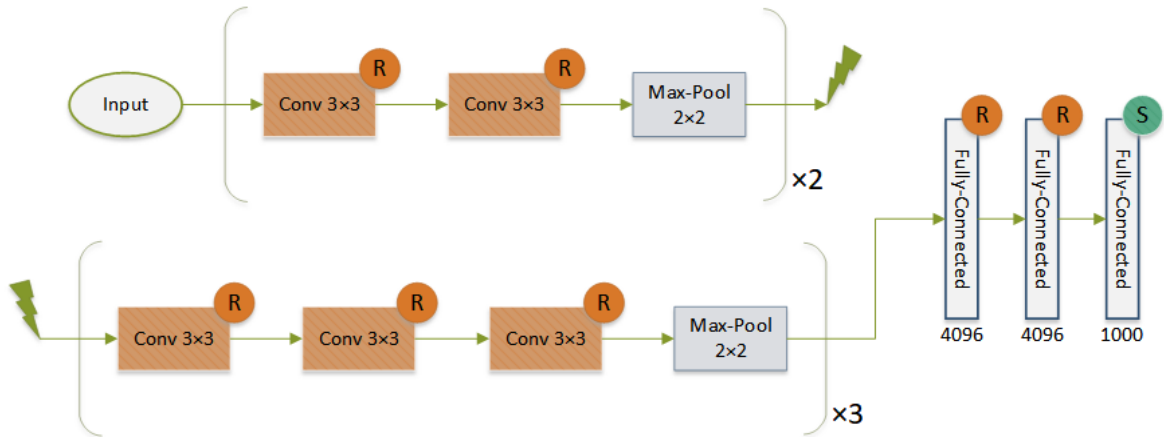

*Figure 9: VGG-16 architecture.*

### 4.3.1.3    GoogleNet:

The architecture of GoogleNet (Inception-v1) proposed by Szegedy et al. [45] uses an Inception module to reduce the number of parameters. The idea of the inception module is that it processes a specific input by multiple parallel convolution layers with different settings. This architecture uses concatenation layers to merge the parallel convolution lines.

The GoogleNet architecture also increased the layers to (22); however, the number of learnable parameters used, (5) million, is 12 times lesser than AlexNet with significantly better accuracy. GoogleNet removed one fully connected layer and replaced it with a (7×7) average pooling layer for an extreme dimensionality reduction.

As shown in figure (10), GoogleNet has parallel lines of convolutions with different filters (1×1), (3×3), and (5×5), followed by a concatenation layer. The max-pooling layer does not reduce the dimensionality since it has a stride value of (1). It just repeats the highest activation value across a (3×3) region. Also, the Inception module uses the (1×1) convolution layers to reduce the number of feature maps and add an extra non-linearity.

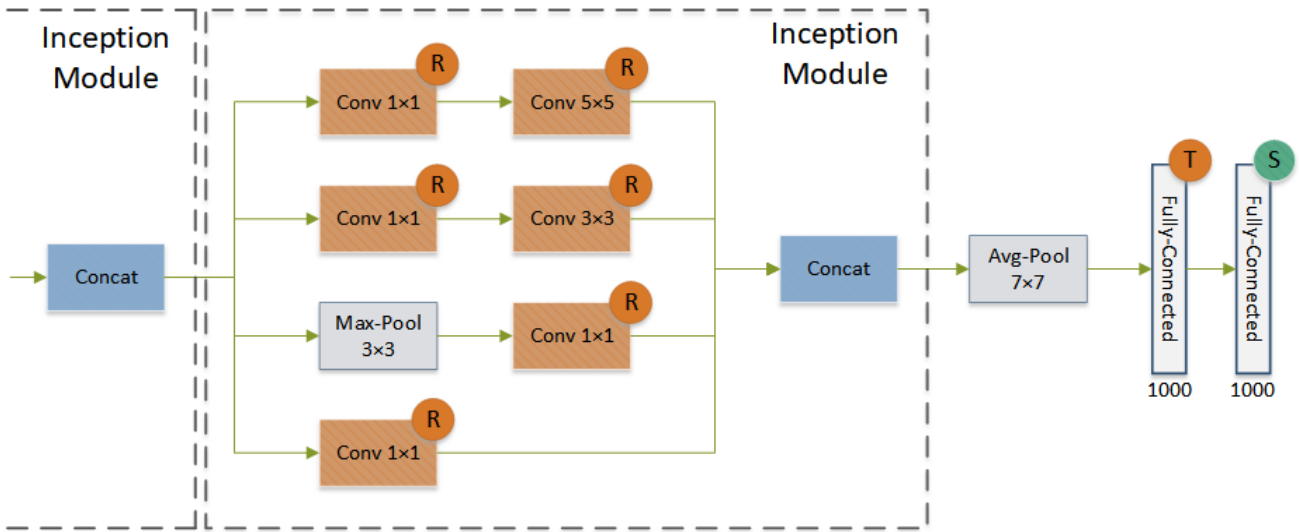

*Figure 5: GoogleNet last Inception module.*

#### 4.3.1.4 Residual Network (ResNet):

As seen in the previously mentioned CNNs, researchers have increased network layers' quantity and deepness. Deeper networks have better performance, but the accuracy gets saturated or even degrades. Accuracy degradation occurs due to the network's inability to train the weights of the first layers. In other words, adding more layers to a suitably deep CNN architecture might have a lower performance.

Residual Network (ResNet), designed by Kaiming He et al. [46] in 2015, introduced a new concept. This new concept explained how deeper models learn through bypass lines called (Skip-Connections). He addressed the deeper networks' problem by using skip-connections to provide a path for the network to train the first layer of the network. Figure (11) illustrate the identity blocks of ResNet. ResNet can have up to (152) layers in total without any degradation. Convolution layers also perform max-pooling operations by setting a stride value of (2).

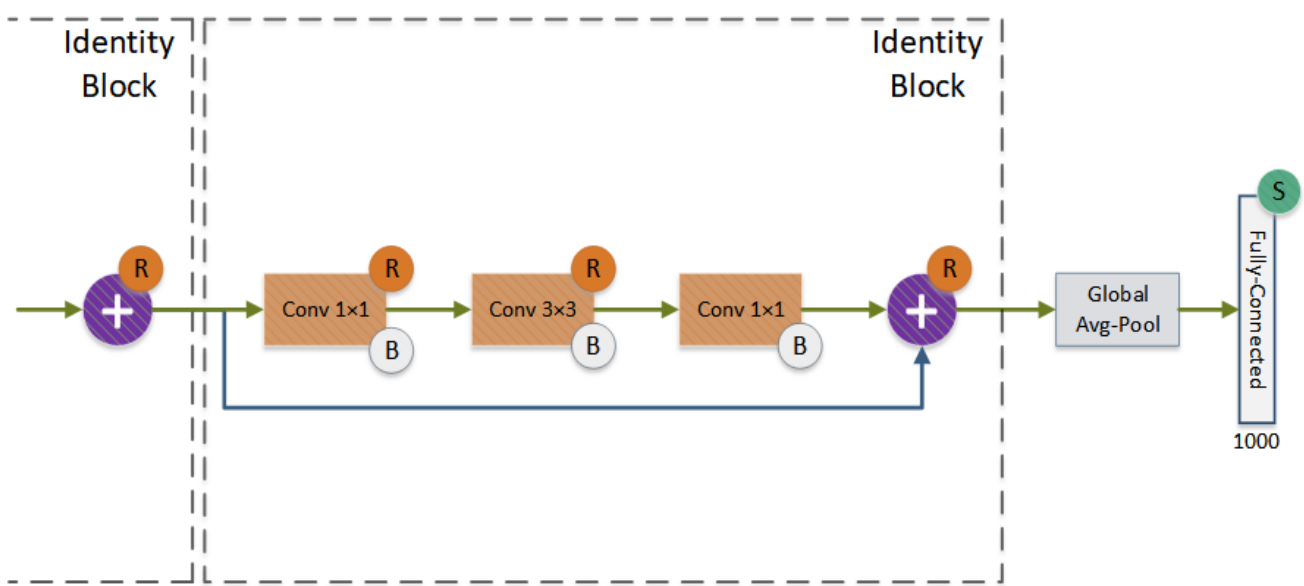

*Figure 6: Last Identity block of ResNet architecture.*

### 4.3.2     CNNs training and Hyperparameters:

The CNN learning process advances by changing the architecture learnable parameters' values iteratively to find the best hypothesis that links the network's input to its desired output. This learning process is called network training. Network training relies on different algorithms that minimize the error rate between the network's actual and desired output. The network training has three stages: feed-forward of input data, calculation and back-propagating the associated error [47] , and learnable parameters adjustment. Stochastic Gradient Descent with Momentum (SGDM) [48] and Cross Entropy Loss function [49] are suitable for this type of classification task training. We trained the network with a (0.001) learning rate. The learnable parameters are regularized with an L2 regularization factor of (0.1) to reduce the network overfitting. CNNs training stops after validation patience of (35) epochs, i.e., the training halts if validation loss did not go any lower after (35) training epochs. To avoid overfitting [50], [51] to the training set, five methods in total were used:

- Data augmentation.
- Reduced number of learnable parameters.
- Dimensionality reduction through global average pooling layer.
- Dropout layer (50% drop).
- L2 regularization term

Despite all this, overfitting can still be present in the model due to information leakage during hyperparameters optimization. As a result, evaluating model performance on test data sets separated from training and validation datasets is the best way to record model performance. The figure below illustrates the overfitting of a network.

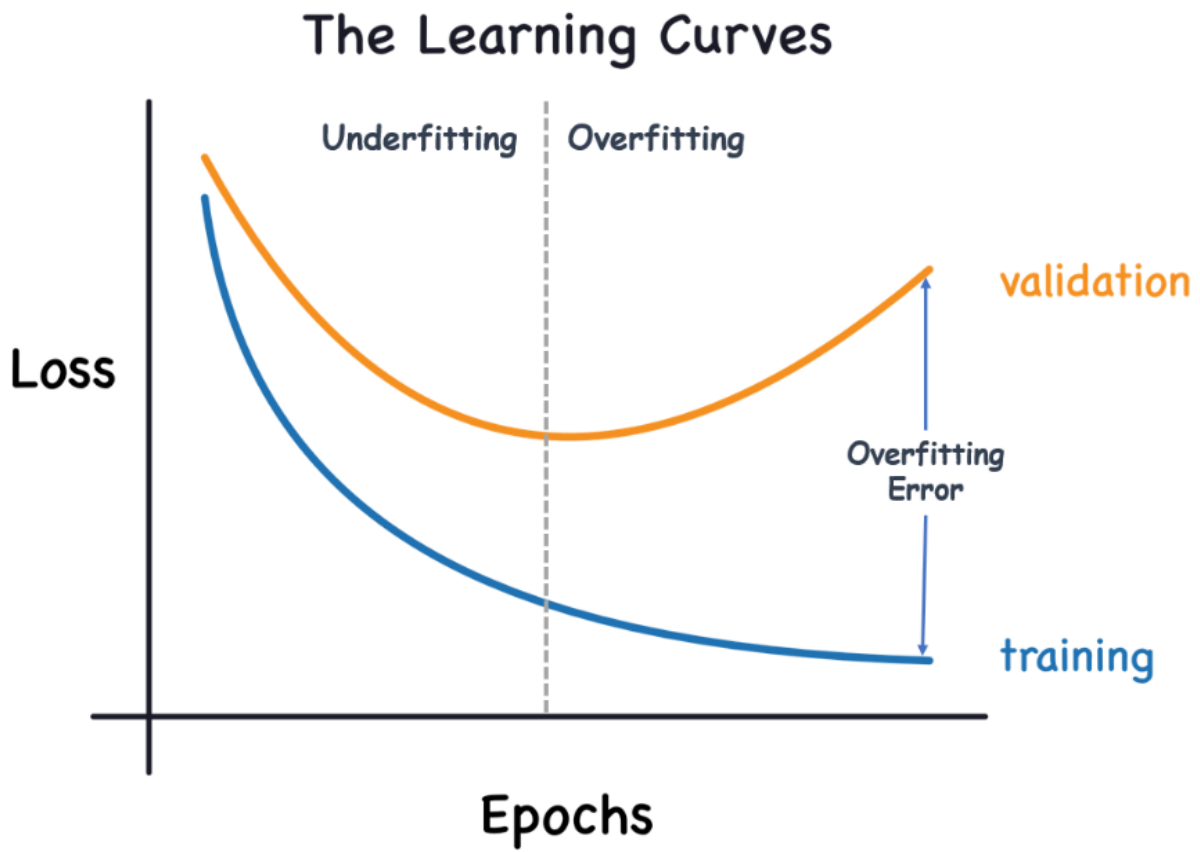

*Figure 7: Overfitting and Underfitting.*

# 5 EXPERIMENTAL RESULT:

This section provides the segmentation evaluation in the first part. The second part delivers the (4) CNN architectures classification results for the proposed method. The second part also illustrates the classification performance of the tissue segmented images against the pre-processed images.

## 5.1 Segmentation Evaluation:

This paper evaluates the accuracy of several segmentation methods segmenting GM, WM, and CSF tissues on the ADNI dataset. The segmentation evaluation relies on computing the overlap between the labeled images and the ground-truth images using pixel Accuracy, Sensitivity, Specificity, Precision, Jaccard, and Dice similarity coefficient. Table (1) shows an evaluation comparison of different segmentation methods.

| method | Accuracy | Sensitivity | Specificity | Precision | Dice | Jaccard |
|--------|----------|-------------|-------------|-----------|------|---------|
| K-means | 92.585% | 89.639% | 94.388% | 89.605% | 89.148% | 80.537% |
| OTSU | 90.174% | 86.410% | 92.644% | 86.525% | 85.597% | 75.073% |
| Fuzzy C-means | 90.686% | 87.012% | 92.973% | 87.262% | 86.351% | 76.191% |
| **K-Means-HMRF** | **96.522%** | **94.943%** | **97.317%** | **94.919%** | **94.920%** | **90.359%** |
| OTSU-HMRF | 95.587% | 93.793% | 96.661% | 93.460% | 93.527% | 87.904% |
| K-means-PSO | 92.588% | 89.639% | 94.388% | 89.605% | 89.148% | 80.537% |
| K-means-GA | 92.590% | 89.647% | 94.392% | 89.610% | 89.155% | 80.549% |

*Table 1: Segmentation methods evaluation.*

The table above proves that the *K*-means clustering technique and HMRF probabilistic model achieved the highest accuracy amongst other segmentation techniques. This duo segmentation methods depend on two features instead of one, the first segmentation feature is the voxel intensity, and the second feature is the voxel neighborhood. *K*-means alone is weak against noise since it does not consider voxel spatial information. *K*-means provides the initial segmentation for the HMRF model depending only on voxel intensity, while HMRF will further optimize the segmentation depending on both voxel intensity and neighborhood feature.

The chart below shows the execution time for each algorithm to perform the segmentation of a single brain MR image. OTSU and K-means are the fastest to complete the segmentation task; (0.140) seconds and (3.094) seconds, respectively. K-means with PSO and K-means with GA required execution times up to (53.232) and (64.933) seconds, respectively, which is highly inefficient due to its lower accuracy.

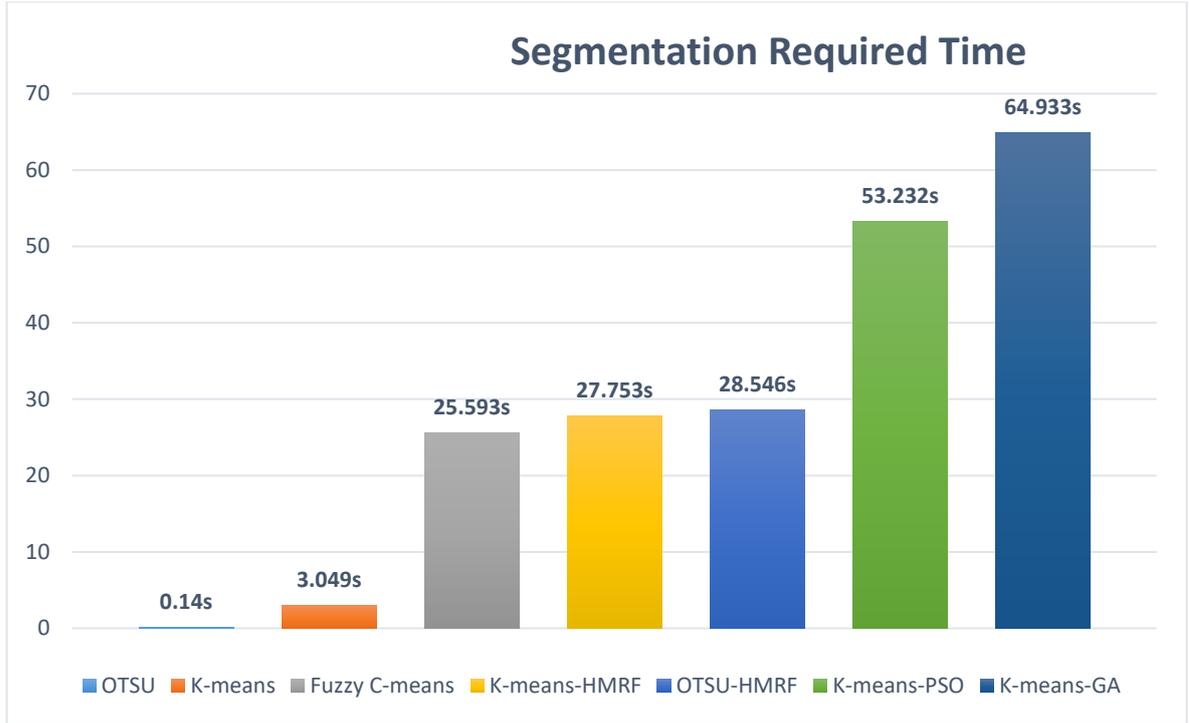

*Figure 8: Required time for different segmentation methods.*

## 5.2 CNN architectures training results :

This work used 3D T1-weighted structural brain MR images downloaded from the ADNI1 dataset to train the CNN architectures. We designed a shallow version of AlexNet, VGGNet, GoogleNet, and ResNet using the Deep Network Designer toolbox with the lowest convolution kernels. Then, we gradually incremented the needed convolution kernels until achieving performance saturation. The training time of each CNN model takes more than one hour; the training is repeated as many times as required to fine-tune the hyperparameters. These trained CNN models represent the feature extraction and classification methods.

The MRI dataset splitting was random to get three groups: 60% training, 20% validation, and 20% testing sets. Data splitting was done based on (5) age groups to ensure age variation between subjects within each groupset. The table (2) represents data splitting for training, validation and test group set.

| | Number of total subjects | AD | CN |
|---|---|---|---|
| *All* | 400 | 200 | 200 |
| *Training Set* | 240 | 120 | 120 |
| *Validation Set* | 80 | 40 | 40 |
| *Testing Set* | 80 | 40 | 40 |

*Table 2: ADNI grouping for training, validation and test group set.*

CNN architecture must go deeper to extract deep features from an image and reduce the overfitting error. The CNN model contains successive convolutional operations to extract deeper features; however, deep CNNs are hard to train due to the vanishing gradient problem. Repeated multiplication operations of convolution layers make the gradients smaller and smaller. This issue is solvable by introducing a batch normalization layer after every convolutional layer.

All training MR images are augmented randomly before entering the CNN at every training epoch to reduce the overfitting to the training dataset. Random augmentation includes:

- [-3 to 3] voxels 3D translation.
- [-5 to 5] degrees 3D rotation.
- [-5 to 5] degrees 3D shearing.
- [95% to 105%] scaling.
- Flipping along the axial view.

The table (3) represents CNN architectures evaluation on the pre-processed images, and tissue segmented images.

| Architecture | Pre-processed images | Tissue-labeled images |
| :---: | :---: | :---: |
| AlexNet | 85.00% | 85.50% |
| GoogleNet | 86.67% | 89.67% |
| VGGNet | 86.67% | 90.33% |
| ResNet | 90.83% | 93.75% |

*Table 3: CNN Classification results for both labeled and original images.*

CNN architectures training was purposed to learn the patterns that discriminate AD subjects from CN subjects. We trained each CNN architecture on both pre-processed images and tissue-labeled images. The overfitting was still present in the models trained on the pre-processed images despite the performed generalization methods. ResNet had the lowest overfitting error with excellent classification accuracy compared to other CNN models because it includes skip-connections that introduce extra non-linearity to the CNN model.

Tissue-labeled images improved classification accuracy upon pre-processed images. Because pre-processed images had an 8-bit value range (0-255), this range increased the complexity of identifying desired features and caused the network to overfit the training data. Tissue labeled images have only (3) data bins (GM, WM, CSF) that prevented the CNN model from identifying features that do not generalize to the testing data. Hence, the CNN architectures trained on the tissue-labeled images had higher classification accuracies when compared to its counterpart.

The only drawback of lowering the image data bins, the training became less stable because of no stepping values between the intensity regions. This issue led to more sensitive convolution operations, and the learnable parameters updates had more impact on the classification result. Solving this required saving checkpoints of the CNN training at the end of every epoch. After the training ends, the CNN training returns the checkpoint with the highest classification accuracy as the trained model.

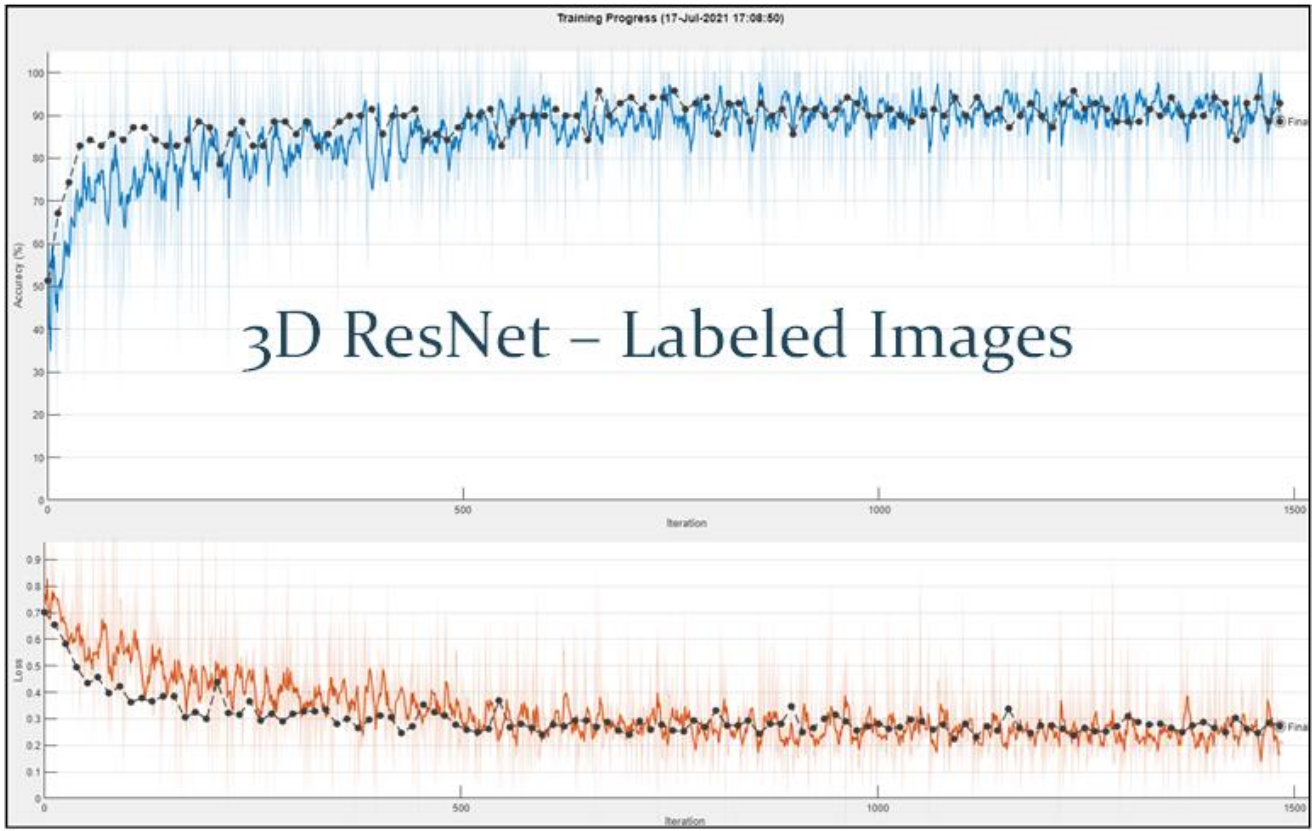

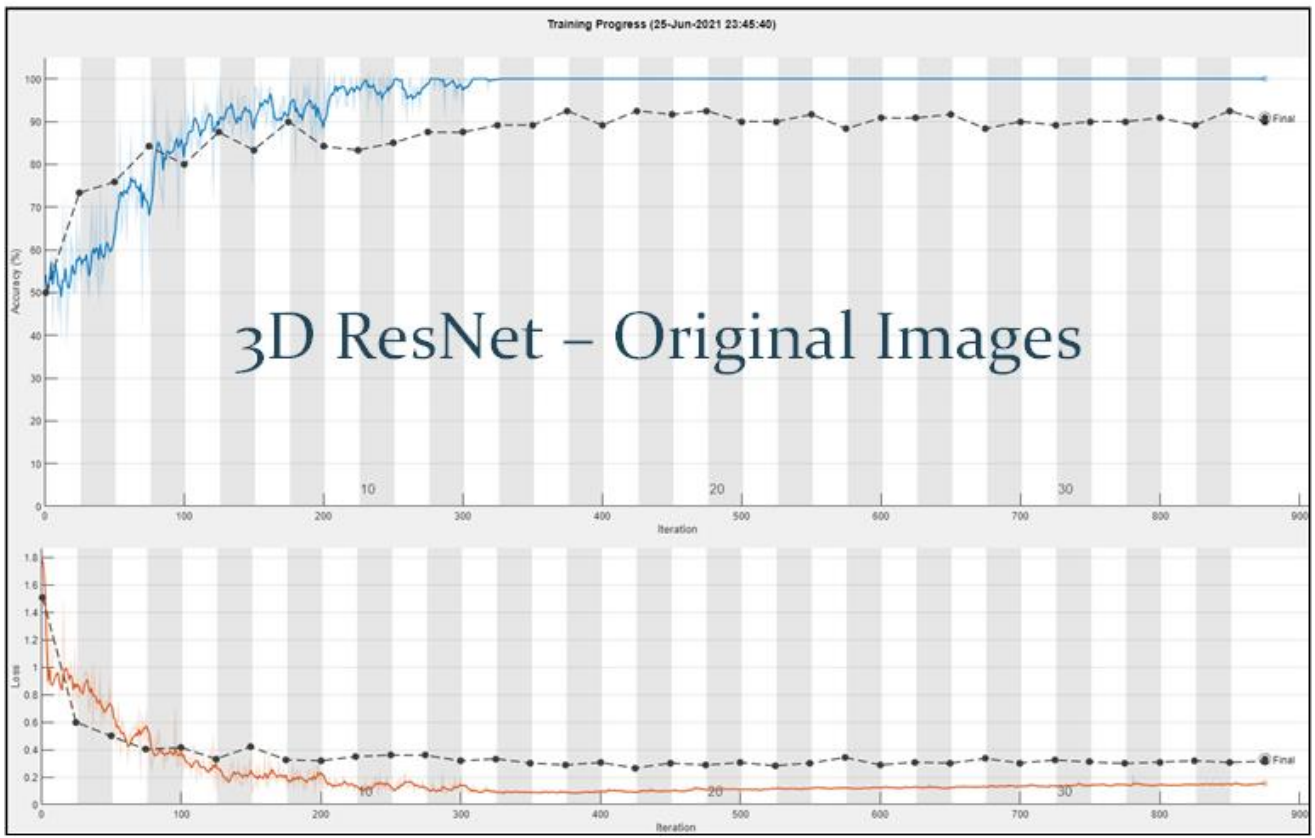

*Figure 9: shows the training vs. validation accuracy and loss for ResNet. Architecture with original images and Tissues label images.*

Figure (14) shows the ResNet training of both types of images. The pre-processed images training suffers from high overfitting error. This overfitting error leads to poor generalization causing lower test data classification results. In the tissue-labeled images, the ResNet CNN training had no overfitting error.

Figure (15) shows that the pre-processed image histogram reveals a high overlap between the intensity regions of the tissues. This overlap causes, in addition to the high amount of data bins, causes high feature extraction complexity. Tissue-labeled images introduce no overlap and reduce data bins to (3) besides the background voxels.

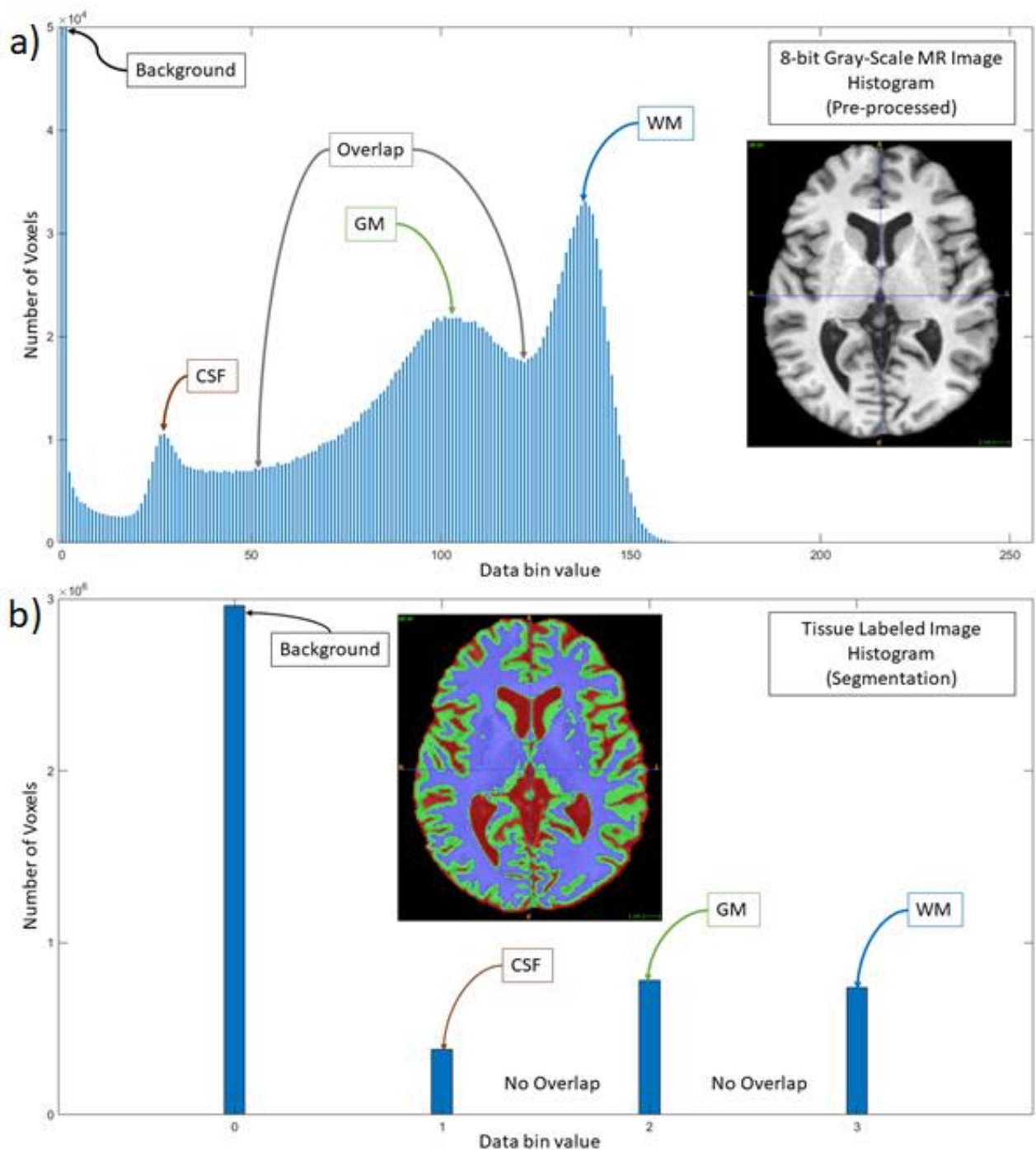

*Figure 10*: *Histogram Difference between a) Regular pre-processed image and b) Segmentation tissue-labeled image.*

# 6 DISCUSSION:

AD is a progressive and degenerative neurological disease. The clinical symptom is that patients fall into dementia in their later years. Due to the increasing cost of nursing for AD, early and accurate diagnosis is vital. DL methods have made significant contributions to AD diagnosis. This paper aims to classify the MR images of the human brain and investigate the influence of image segmentation on the classification process. In this study, two proposed approaches exploited the most commonly acquired imaging technique; anatomical MRI of the brain. The first approach uses the pre-processed MR images, and the second one uses the post-processed tissue-labeled images as inputs to the feature extraction and classification process. The work aimed to accomplish three goals:

- Pathological AD-related patterns recognition and AD vs. CN binary classification.
- Investigate the outcome of image segmentation on (4) CNN architectures' classification performance.
- Outperform the state-of-art methods.

We proposed a classification software pipeline that used the brain MR Image as input. This pipeline employs the best-performing CNN model among (4) shallow architectures. We trained These CNN architectures on two distinct inputs; pre-processed MR images and tissue-labeled images. Our experimental result illustrates that the labeled images achieved better accuracy when compared to pre-processed images within the (4) CNN architectures. Table (3) shows the best-performing CNN model compared with the state-of-the-art method.

- ❖ Several studies employed one or more 2D slices of the 3D brain MRI. They achieved low classification accuracy because of information loss when neglecting other brain regions not included in the chosen 2D slices.
- ❖ Several studies employed all 2D slices of the 3D brain MRI. Although they achieved better classification results, 2D interpretation of the 3D world also suffers from information loss, given that MRI scans are 3D images, and a spatial relationship exists among 2D image slices.
- ❖ Several studies proposed 3D DL architectures to classify the 3D brain MR images; these studies suffered from over-fitting due to high dimensionality.
- ❖ In the proposed method, we used the 3D tissue-labeled MR images. Such low data bins images reduced the overfitting and increased the learnable parameter ability to learn the dominant features leading to improved accuracy.

| References | Approach | AD vs. CN |
|---|---|---|
| Yagis et al. [24] 2020 | 3D CNN | 73.40% |
| Pan et al. [22] 2020 | CNN & EL | 84.00% |
| Li et al. [26] 2018 | Dense Nets | 89.50% |
| Cheng et al. [27] 2017 | Multi-Deep CNN | 87.15% |
| Korolev et al. [28] 2017 | ResNet | 80.00% |
| Zhang et al. [29] 2021 | 3D ResAttNet | 91.30% |
| Proposed method | 3D ResNet-20 | 93.75% |

*Table 4: Comparison with state-of-the-art methods.*

All studies in the table above relied on regular MRI images; the proposed DL model relied on tissue-labeled images. Our investigation proved that tissue-labeled images achieved higher performance when compared to unsegmented images. The proposed DL method achieved (93.75%) accuracy for AD vs. CN classification by ResNet architecture, which is higher than reported DL techniques.

# 7 CONCLUSION:

AD diagnosis is a challenging task which many authors are focusing on nowadays. They had developed many computer-aided diagnosis (CAD) systems to perform the diagnosis of AD. This paper explains an automatic AD diagnosis system based on deep learning and the 3D brain structural MRI scans provided by Alzheimer's disease Neuroimaging Initiative.

Brain MRI scans undergo many processing steps to enhance the visual appearance and reduce the complexity of proceeding steps. The proposed method employs a combo of K-means clustering and Hidden Random Markov Random Field for the three brain tissues segmentation: White Matter, Gray Matter, and Cerebrospinal Fluids. This combo segmentation method achieved a high accuracy using two image characteristics instead of one, voxel intensity and voxel neighborhood. We performed features extraction and classification using deep neural network architectures. The main advantage of Neural Networks compared to other classic methods; Neural Networks can automatically recognize patterns from raw data without expert supervision.

The experimental results demonstrated that the proposed model enhances brain 3D MR images using the segmented tissues. These tissue-labeled images had a lower feature extraction complexity when compared to regular MR images. Such methodology proved to be another overfitting reduction scheme that adds another novelty to this paper. This method gave an accurate prediction of the AD vs. CN binary classification. The proposed model achieved results that exceeded other DL state-of-the-art methods.

In the future, we aim to enhance the MR images using image segmentation instead of using tissue-labeled images. This idea may reduce the instability of the network training by adding more stepping values between the intensity regions. Another idea is to utilize other modalities to enhance these images and boost classification performance. We also intend to construct an ensemble deep learning architecture that uses several inputs such as Regions of Interests that carry more features and facilitate the feature extraction process.